\documentclass[twocolumn,showpacs,floatfix]{revtex4}%
\usepackage{graphicx}%
\usepackage{amsmath}%
\usepackage{amsfonts}%
\usepackage{amssymb}
\usepackage{bm}

\def\e{{\epsilon}}
\def\k{{ {\bm k} }}

\def\w{{\omega}}

\allowdisplaybreaks[4]

\begin{document}
\title{
Study of phase diagram and superconducting states in LaFeAsO$_{1-x}$H$_x$ \\
based on the multiorbital extended Hubbard model
}
\author{Y. Yamakawa$^{1}$, S. \textsc{Onari}$^{2}$,
H. \textsc{Kontani}$^{1}$,
N. \textsc{Fujiwara}$^{3}$,
S. \textsc{Iimura}$^{4}$, and
H. \textsc{Hosono}$^{4}$
}

\date{\today }

\begin{abstract}

To understand the recently established unique magnetic and 
superconducting phase diagram of LaFeAsO$_{1-x}$H$_x$, 
we analyze the realistic multiorbital tight-binding model
for $x=0\sim0.4$ beyond the rigid band approximation.
Both the spin and orbital susceptibilities are calculated
in the presence of the Coulomb and charge quadrupole interactions.
It is found that both orbital and spin fluctuations 
strongly develop at both $x\sim0$ and $0.4$,
due to the strong violation of the rigid band picture in LaFeAsO$_{1-x}$H$_x$.
Based on this result,
we discuss the experimental phase diagram,
especially the double-dome superconducting phase.
Moreover, we show that the quadrupole interaction is 
effectively produced by the vertex correction due to Coulomb interaction,
resulting in the mutual development of spin and orbital fluctuations.

\end{abstract}

\address{
$^1$ Department of Physics, Nagoya University,
Furo-cho, Nagoya 464-8602, Japan. 
\\
$^2$ Department of Applied Physics, Nagoya University,
Furo-cho, Nagoya 464-8602, Japan. 
\\
$^3$ Graduate School of Human and Environmental Studies, Kyoto University,
Yoshida-Nihonmatsu-cho, Sakyo-ku, Kyoto 606-8501, Japan.
\\
$^4$ Materials and Structures Laboratory, Tokyo Institute of Technology, 4259 Nagatsuta-cho, Midori-ku, Yokohama 226-8503, Japan.
}
 
\pacs{74.70.Xa, 74.20.-z, 74.25.Dw}

\sloppy

\maketitle

%%%%%%%%%%%%%%%%%%
%Introduction
%%%%%%%%%%%%%%%%%%
Since the discovery of high-$T_{\rm c}$ superconductivity 
in Fe-based superconductors
\cite{Kamihara}, its pairing mechanism has been studied very intensively. 
In unconventional superconductors,
the phase-diagram in the normal state 
gives us very important hints to understand the 
mechanism of superconductivity.
In many heavy fermion superconductors, 
%add
for example,
the superconducting (SC) phase is next to the magnetic ordered phase, 
indicating the occurrence of spin-fluctuation mediated superconductivity.
In Fe-based superconductors, in contrast,
the ferro-orbital order occurs in the orthorhombic phase 
\cite{Shimojima,ARPES-Shen}, 
and the structure or orbital instabilities are realized in the normal state
\cite{Yoshizawa,Goto},
in addition to the magnetic instability.
Based on this fact, 
both the spin-fluctuation mediated $s_\pm$-wave state 
\cite{Mazin,Kuroki,Hirschfeld,Chubukov}
and the orbital-fluctuation mediated $s_{++}$-wave state 
\cite{Kontani-RPA,Kontani-SSC,Onari-SCVC}
have been discussed.
The former (latter) SC gap with (without) sign reversal
is induced by repulsive (attractive) interaction between
electron-like and hole-like Fermi surfaces (FSs).

Recently, very unique phase diagram of H-doped LaFeAsO,
LaFeAsO$_{1-x}$H$_x$,
%shows very unique phase diagram shown in Fig. of
%Ref. \cite{Hosono-Hdoped}:
is determined in Ref. \cite{Hosono-Hdoped}:
The structure and magnetic transitions
in mother compound are replaced with the SC phase at $x\sim0.03$,
and interesting double-dome structure of $T_{\rm c}$ is 
obtained between $x=0.03$ and $\lesssim 0.5$.
This fact may indicate that the two different pairing mechanisms 
are involved in under- and over-doped regions.
The maximum $T_{\rm c}$ of the first (second) dome is about
25 K (40 K)  at $x\sim0.1$ ($x\sim0.35$).
For $x>0.4$,  recent NMR measurement \cite{Fujiwara} detected 
the incommensurate magnetic order, in addition to 
the highly anisotropic electric field gradient
that indicates the occurrence of the non-magnetic orbital order.

In LaFeAsO$_{1-x}$H$_x$, the electron filling per Fe is $n=6+x$
since each H-dopant becomes H$^{-}$ ion,
and therefore the electronic states are expected to be similar to 
those of LaFeAsO$_{1-x}$F$_x$ \cite{Hosono-Hdoped}.
Based on this fact, the band structure and FSs
for $x=0\sim0.4$ had been derived from the local-density-approximation (LDA)
band calculation using the virtual crystal approximation
in Ref. \cite{Hosono-Hdoped}.
It is found that the rigid band picture is no more valid,
since the band structure is strongly modified with increasing H-doping.
%for example, the bandwidth is remarkablyrenormalized with F-doping.
%Also, the level-splitting between $xz/yz$-orbital and $xy$-orbital
%becomes very small for $x\sim0.4$,
%which is a favorable condition for realizing the strong orbital fluctuations
% \cite{Hosono-Hdoped}.
The derived realistic band structure now
enables us to perform quantitative theoretical study of the 
pairing mechanism of 1111 systems.

In this paper, we study the electronic and SC states
in LaFeAsO$_{1-x}$H$_x$ for $x=0\sim0.4$, by constructing the 
realistic multiorbital models beyond the rigid band approximation.
%given by the first principle study.
Using the random-phase-approximation (RPA),
both the spin and orbital susceptibilities are calculated 
in the presence of the Coulomb interaction $U$
and charge quadrupole interaction $g$. 
Assuming monotonic $x$-dependencies of these interactions,
strong spin and orbital fluctuations
are obtained for both $x\sim 0$ and $x\sim0.4$.
%For $x\gtrsim0.4$, incommensurate magnetic order is expected.
Based on this result, the origin of the double-dome structure of 
$T_{\rm c}$ in LaFeAsO$_{1-x}$H$_x$ is discussed,
by applying both the orbital-fluctuation mediated $s_{++}$-wave scenario
and spin-fluctuation mediated $s_\pm$-wave one.
We discuss that $g$ is effectively induced by the vertex correction (VC) 
of the Coulomb interaction beyond the RPA.
%, which is the many-body interaction beyond the RPA.
%By going beyond the rigid-band approximation,

In LaFeAsO$_{1-x}$F$_x$, the spin fluctuations 
observed by NMR and neutron inelastic scattering
are very small in slightly over-doped compounds ($x\sim0.08$).
In the over-doped compound with $x=0.14$,
$T_{\rm c}$ increases from 20K to 43K by applying 3.7GPa,
irrespective that $1/T_1T$ remains very small 
independently of the pressure \cite{Nakano}.
These facts indicate the weak correlation between 
$T_{\rm c}$ and spin fluctuations.
Moreover, impurity effect on $T_{\rm c}$ is 
very small in both 1111 \cite{Sato-imp,Li-1111}
and 122 \cite{Nakajima,Li} compounds,
indicating the realization of the $s_{++}$-wave state
\cite{Onari-imp,Yamakawa-imp}
or $s_{\pm}\rightarrow s_{++}$ crossover \cite{Onari-imp,Efremov}.
Unfortunately, these experiments on LaFeAsO$_{1-x}$H$_x$
have not been performed.

%%%%%%%%%%%%%%%%%%%%%%%%%%%%%%%%%
\begin{figure}[!htb]
\includegraphics[width=0.9\linewidth]{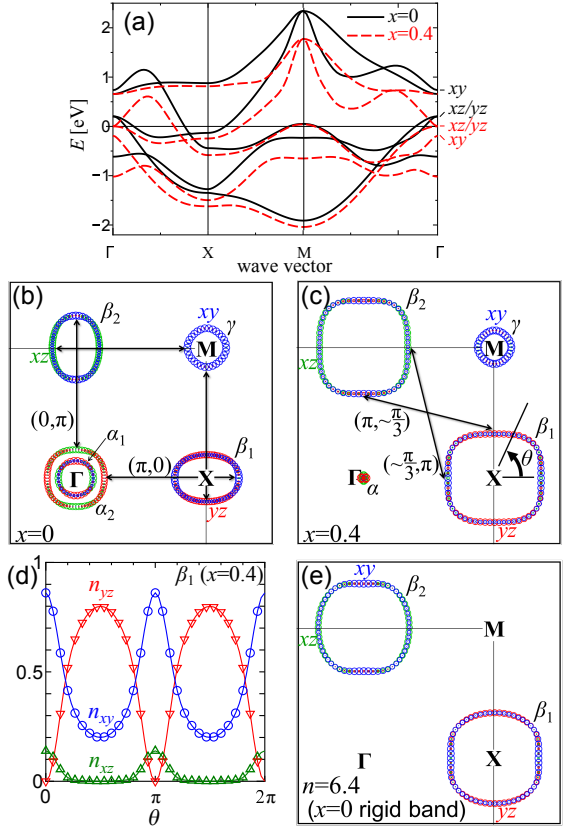}
\caption{
(Color online)
(a) Band structures for $x=0$ (solid line) and
 $x=0.4$ (dashed line) in the present model. 
(b) FSs for $x=0$  and (c) FSs for $x=0.4$.
The weights of $xz$-, $yz$- and $xy$-orbital on the Fermi surfaces
are represented by diameter of green-, red-, and blue-circles, respectively.
(d) Weight of orbitals on the Fermi surface around X point for $x=0.4$ 
as a function of azimuthal angle $\theta$. 
(e) FSs for $n=6.4$ given by the rigid band model for LaFeAsO.
}
\label{fig:Fermi}
\end{figure}
%%%%%%%%%%%%%%%%%%%%%%%%%%%%%%%%%

%%%%%%%%%%%%%%%%%%%%%%%%%%%%%%%%%%%%%%%%%%%%%%%%%%%%%%%%%%%%%%%%%

%Experimentally, the electronic states of 
%LaFeAsO$_{1-x}$H$_x$ are expected to be
%similar to those of LaFeAsO$_{1-x}$F$_x$ 
%since H$^{-1}$ ion acts as F$^{-1}$ ion \cite{Hosono-Hdoped}.
Here, we perform the LDA band calculation
for LaFeAsO$_{1-x}$F$_x$ for $x=0\sim0.4$
using WIEN2K code, with virtual crystal
approximation, in which the oxygen sites are
substituted for virtual atoms with a fractional nuclear charge.
In the full-potential LDA, the virtual crystal approximation 
works when the core electrons of the original atom
and those of the substitutional atom are the same.
We use the  crystal structures of LaFeAsO$_{1-x}$H$_x$ reported in 
Refs. \cite{Hosono-Crystal}.
Next, we derive the two-dimensional five-orbital tight-binding models 
for each $x$ 
using WANNIER90 code and WIEN2WANNIER interface \cite{win2wannier}:
$\hat{H}^{0}= \sum_{\bm{k} l m \sigma}
h^{l,m}_{\bm{k}} c^{\dagger}_{\bm{k}l\sigma} c_{\bm{k}m\sigma}$,
%
%
%\begin{equation}
%\hat{H}^{0}= \sum_{\bm{k} l m \sigma}
%h^{l,m}_{\bm{k}} c^{\dagger}_{\bm{k}l\sigma} c_{\bm{k}m\sigma}
%\end{equation}
%
where $l,m = 1-5$ represent the $d$ orbitals with the order
$3z^2-r^2$, $xz$, $yz$, $xy$, and $x^2-y^2$:
Here, we set $x$ and $y$ axes parallel to the nearest Fe-Fe bonds.

Figure \ref{fig:Fermi} (a) shows the band structures of LaFeAsO$_{1-x}$F$_x$
in the present model. 
It is obvious that band
structure for $x=0.4$ cannot be reproduced by the rigid band shift from
that for $x=0$.
%We stress that level splitting between $xy$-orbital and $xz/yz$-orbital
%at $\Gamma$ point becomes smaller for
%$x=0.4$ than that for $x=0$.
%Thus, orbital fluctuations are enhanced due to the degeneracy of orbitals
%for $x=0.4$.
The corresponding FSs are shown in Fig. \ref{fig:Fermi} (b) and (c).
Here, $\beta_1$ and $\beta_2$ are the electron-pockets, and 
$\alpha_1$, $\alpha_2$ and $\gamma$ are the hole-pockets,
both of which are composed of the three $xz$-, $yz$- and $xy$-orbitals:
The orbital character of the electron-pocket for $x=0.4$
is shown in Fig. \ref{fig:Fermi} (d).
In the case of $x=0$, the electron-hole (e-h) FS nesting 
with the nesting vector $\bm{Q}=(\pi,0),(0,\pi)$
is the most important.
%electron-hole (e-h) nestings
%between X $\Gamma$, M and X points
%$\bm{Q}=(\pi,0),(0,\pi)$ are dominant because size of FSs are almost same.
%Focusing on the orbital dependence, we see that these nestings consist
%of both inter-orbital and
%intra-orbital nesting of 
On the other hand, in the case of $x=0.4$, electron-electron (e-e) FS nesting
$\bm{Q}\sim(\pi,\pi/3),(\pi/3,\pi)$ is more important 
since the hole pockets become very small.
In both $x=0$ and $x=0.4$,
both the intra-orbital nesting (mainly $xy$) and inter-orbital nesting
(between $xz/yz$ and $xy$) are important.
Then, the former (latter) nesting 
gives rise to the strong spin (orbital) fluctuations,
as discussed in Ref. \cite{Kontani-RPA}.

Fig.  \ref{fig:Fermi} (e) shows the FSs for $n=6.4$
given by the model parameters for $x=0$ (LaFeAsO).
In this ``rigid-band approximation'', the hole pockets 
disappear, and the e-e FS nesting is worse
since the shape of the electron-pockets are more rounded.

Next, we explain the interaction term.
We introduce both the Coulomb interaction ($U$, $U'$, $J$, $J'$)
and quadrupole interaction ($g$).
%The latter are induced by the e-ph interaction
%due to Fe ion oscillations as follows,\cite{Kontani-AL}
The latter interaction is 
\begin{equation}
V_{\rm{quad}}=-  g(\w_l) \sum_i^{\mathrm{site}} \left( 
{\hat O}^i_{xz} \!\cdot\!{\hat O}^i_{xz} + 
{\hat O}^i_{yz} \!\cdot\!{\hat O}^i_{yz} + 
{\hat O}^i_{xy} \!\cdot\!{\hat O}^i_{xy}\right) ,
 \label{eqn:Hint}
\end{equation}
where  $\omega_l=2l\pi T$ and $g(\w_l)=g \cdot \w_{\rm c}^2/(\w_l^2+\w_{\rm c}^2)$:
$g=g(0)$ is the quadrupole interaction at $\omega_l=0$, 
and $\omega_{\rm c}$ is the cutoff energy.
$\hat{O}_{\Gamma}$ is the quadrupole operator \cite{Kontani-RPA}, 
which has many non-zero off-diagonal elements.
%In addition, the AL-type vertex correction (two-orbiton/magnon process) due to Coulomb interaction
%causes large effective interaction $g_1$ \cite{OnariVC}.
By introducing small $g \ (\sim 0.2{\rm eV})$, 
strong $O_{xz,yz}$-type antiferro-quadrupole fluctuations
are caused by the good inter-orbital nesting,
as explained in Ref. \cite{Kontani-RPA}.

Now, we perform the RPA for the present model at $T=0.02$eV,
by assuming that $J=J'$ and $U=U'+2J$, and fix the ratio $J/U=1/6$.
We use $64 \times 64 \bm{k}$ meshes and 512 Matsubara frequencies.
We set the unit of energy as eV hereafter.
The spin (orbital) susceptibility in the RPA is given by %\cite{Takimoto}
\begin{eqnarray}
\hat{\chi}^{\mathrm{s(c)}} \left( q \right) = \hat{\chi}^{0} \left( q \right) 
\left[ \hat{1} - \hat{\Gamma}^{\mathrm{s(c)}}(\omega_l) \hat{\chi}^{0} \left( q \right) 
\right]^{-1}, \label{eqn:RPA}
\end{eqnarray}
%\hat{\chi}^{\mathrm{c}} \left( q \right) = \frac{\hat{\chi}^{0} \left( q \right)}{\hat{1} - \hat{\Gamma}^{\mathrm{c}} (\omega_l) \hat{\chi}^{0} \left( q \right)},
%
%\end{gather}
where $q=({\bm q},\w_l)$, and
$\hat{\Gamma}^{s(c)}$ is the interaction matrix for the spin (charge) channel 
composed of $U,U',J,J'$ and $g(\omega_l)$ \cite{Kontani-RPA}.
$\hat{\chi}^{0} \left( q \right)=
- \frac{T}{N} \sum_k G_{lm} \left( k+ q \right) G_{m' l'} \left( k \right)$
is the irreducible susceptibility, where
$\hat{G} ( k ) = [ i \epsilon_n + \mu - \hat{h}_{\bm{k}} ]^{-1}$
is the bare Green function, and $\epsilon_n = (2n + 1) \pi T$.
%where $\mu$ is the chemical potential.
The magnetic (orbital) order is realized 
when the spin (charge) Stoner factor $\alpha_{\mathrm{s}(\mathrm{c})}$,
which is the maximum eigenvalue of $\hat{\Gamma}^{\mathrm{s}(\mathrm{c})} \hat{\chi}^{(0)} ( \bm{q} , 0)$, is unity.

%In the RPA, the enhancement of $\hat{\chi}^{\mathrm{s}}$
%is mainly caused by the intra-orbital Coulomb interaction $U$,
%using the ``intra-orbital nesting'' of the FSs.
%On the other hand, the enhancement of $\hat{\chi}^{\mathrm{c}}$ 
%in the present model
%is caused by the quadrupole interaction in Eq. (\ref{eqn:Hint}),
%utilizing the ``inter-orbital nesting'' of the FSs.

\begin{figure}[!htb]
\includegraphics[width=0.9\linewidth]{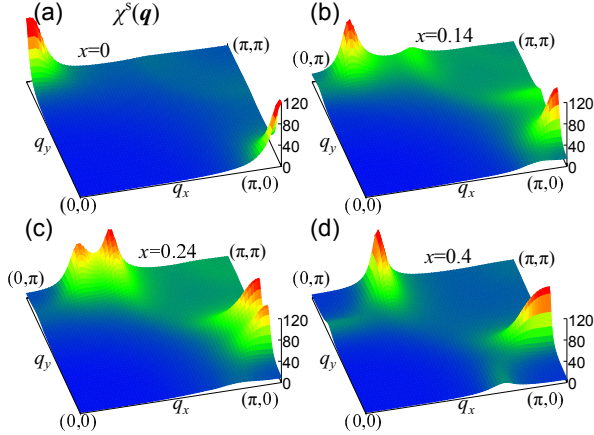}
\caption{
(Color online)
$\bm{q}$-dependence of $\chi^s({\bm q})$ 
with $\alpha_s=0.98$ for (a) $x=0$, (b) $x=0.14$,
 (c) $x=0.24$ and (d) $x=0.4$.
}
\label{fig:chis}
\end{figure}

Now, we study the development of the spin susceptibility by $U$,
by putting $g=0$.
Figure \ref{fig:chis} shows the total spin susceptibility 
$\chi^s({\bm q},\omega=0) \equiv \sum_{l,m}\chi^s_{ll,mm}({\bm q})$ 
for $x=0,0.14,0.24,0.4$, by choosing the $U$ to realize $\alpha_s=0.98$.
For (a) $x=0$,  $\chi^s({\bm q})$ has commensurate peaks at 
${\bm q}=(\pi,0)$ and $(0,\pi)$ due to the e-h FS nesting.
These peaks change to incommensurate for (b) $x=0.14$,
reflecting the size imbalance between electron- and hole-pockets.
As increasing the doping further, 
the e-e FS nesting and e-h FS nesting become comparable.
Because of the fact, $\chi^s({\bm q})$ for (c) $x=0.24$
shows the double-peak structure.
For (d) $x=0.4$,  $\chi^s({\bm q})$ shows the 
incommensurate peak structure due to the e-e FS nesting only.

%%%%%%%%%%%%%%%%%%%%%%%%%%%%%%%%%
%%%%%%%%%%%%%%%%%%%%%%%%%%%%%%%%%

\begin{figure}[!htb]
\includegraphics[width=0.99\linewidth]{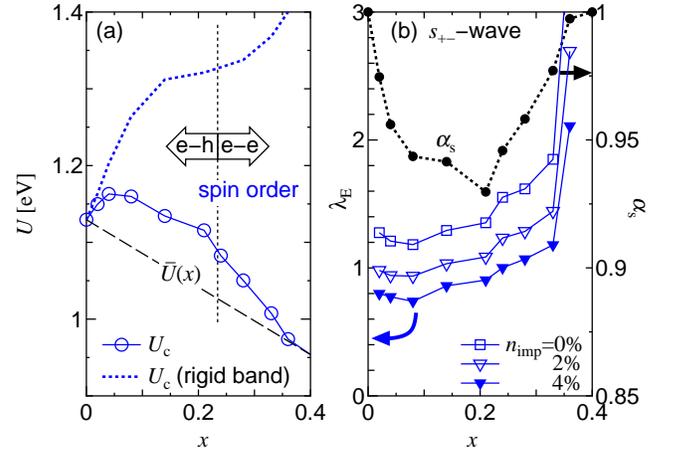}
\caption{
(Color online)
(a) $U_c$ in present models (solid line) and that in the rigid band model
 (dotted line) against $x$ for $g=0$. ${\bar U}(x)$ is determined by liner
 interpolation between $U_c$ for $x=0$ and that for $x=0.4$.
(b) Obtained $\lambda_{E}$ for the $s_{\pm}$-wave state
 with $n_{\rm imp}=0\%,2\%,4\%$.
 The $x$-dependence of $\alpha_s$ is also shown.
}
\label{fig:Spm}
\end{figure}

%Disscussion of $U_{\rm eff}$, $g_{\rm eff}$.
Figure \ref{fig:Spm} (a) shows the $x$-dependence of $U_c$, which is 
the critical value of $U$ for the spin order given by the
condition $\alpha_{s}=1$.
We stress that $U_c$ in the present model is much smaller than that in 
the rigid band model, reflecting the good e-h (e-e) FS nesting for
$x<0.24$ ($x>0.24$) in the present model.  
Moreover, $U_c$ takes the maximum value at $x\approx 0.1$, 
and monotonically decreases by departing from $x\approx0.1$. 
%This fact is very favorable to reproduce the experimental spin and orbital order in the phase diagram.
For this reason, we can explain the magnetic orders at $x\sim0$ and $0.4$,
by assuming a simple monotonic $x$-dependence of the interaction:
Here, we introduce ${\bar U}(x)$ by the liner interpolation between 
$U_c$ for $x=0$ and that for $x=0.4$, as shown in Fig. \ref{fig:Spm} (a).
%In fact, $U_{\rm eff}(x)$ would moderately decrease with increasing $x$,
%because the Kanamori theory predicts 
%$U_{\rm eff}(x)\sim U[1 + \beta UN(x)]^{-1}$, where $\beta\sim1$ and 
%$N(x)$ is the density 
%of states at the Fermi level, and it slightly increases with
%increasing $x$ in the present model\cite{Hosono-Hdoped}.
%Then, we can explain the emergence of spin and orbital ordered states at 
%$x\sim0$ and $x\sim0.4$ within the RPA,
%by introducing {\it monotonic $x$-dependences of $U_{\rm eff}$ and $g_{\rm eff}$}.
The $x$-dependence of ${\bar U}(x)$ might be explained by the change in
the Kanamori screening, since the density of states at the Fermi level,
$N(0)$, increases by $30\%$, by changing $x$ from 0 to 0.4.

%the change in the $d$-electron Wannier functions  \cite{Miyake},
%or the underestimation of the bandwidth 
%given by the first-principle study for $x\sim0.4$.

%%%%%%%%%%%%%%%%%%%%%%%%%%%%%%%%%
%%%%%%%%%%%%%%%%%%%%%%%%%%%%%%%%%

%%%%%%%%%%%%%%%%%%%%%%%%%%%%%%%%%
\begin{figure}[!htb]
\includegraphics[width=0.99\linewidth]{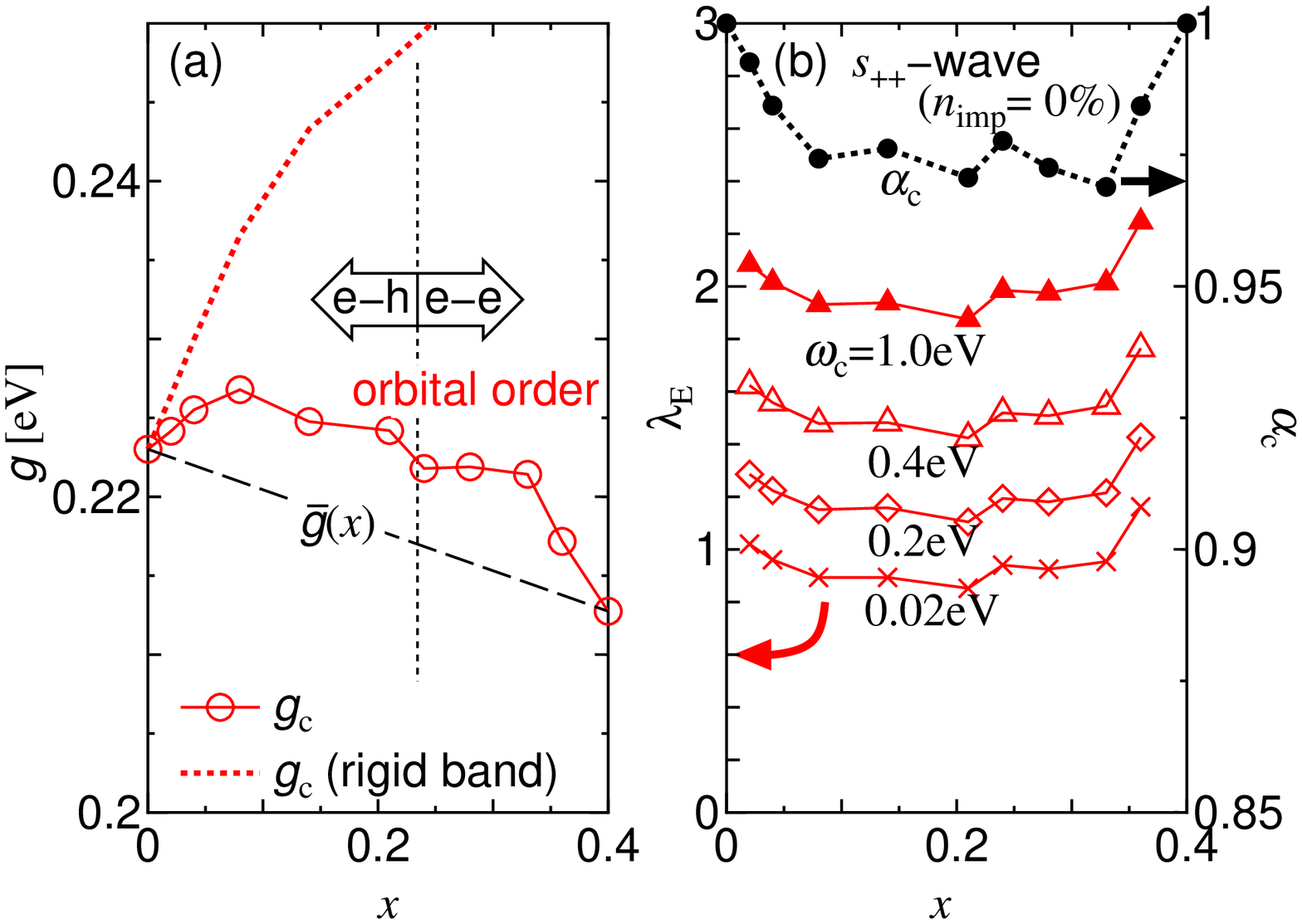}
\caption{
(Color online)
(a) $g_c$ in present models (solid line) and that in the rigid band model
 (dotted line) against $x$ for $U=0$. 
${\bar g}(x)$ is determined by liner
 interpolation between $g_c$ for $x=0$ and that for $x=0.4$.
(b) Obtained $\lambda_{E}$ for the $s_{++}$-wave state
 with $\omega_{\rm c}=0.02\sim1$ ($n_{\rm imp}=0$).
 The $x$-dependence of $\alpha_c$ is also shown.
}
\label{fig:Spp}
\end{figure}
%%%%%%%%%%%%%%%%%%%%%%%%%%%%%%%%%

We also analyze the SC state using the linearized Eliashberg equation:
\begin{eqnarray}
\lambda_{\rm E}\Delta_{ll'}(k)&=-&\frac{T}{N}\sum_{k',m_i}W_{lm_1,m_4l'}(k-k')
G'_{m_1m_2}(k')\nonumber\\
&\times&\Delta_{m_2m_3}(k')
G'_{m_4m_3}(-k')
+\delta\Sigma^a_{ll'}(\epsilon_n),
\label{eqn:Eliash} 
\end{eqnarray}
where $\Delta_{ll'}(k)$ is the gap function, and
$\lambda_{\rm E}$ is the eigenvalue that reaches unity at $T=T_{\rm c}$.
When $T$ is fixed, the larger value of $\lambda_{\rm E}$ 
would correspond to the higher $T_{\rm c}$.
Here, ${\hat W}$ is the pairing interaction given by the RPA, 
$(\hat{G'})^{-1}=(\hat{G})^{-1}-\delta\hat{\Sigma}^n$ 
is the normal Green function with impurity-induced 
normal self-energy $\delta\hat{\Sigma}^n$, and 
$\delta{\hat \Sigma}^a$ is the impurity-induced anomalous self-energy.
The expressions are given in Ref. \cite{Kontani-RPA}.
Hereafter, we put the orbital-diagonal on-site impurity potential as $I=1$.
Figure \ref{fig:Spm} (b) shows the obtained $\lambda_{\rm E}$ as a
function of $x$.
We see that $\lambda_{\rm E}$ has two peaks at both $x=0.04$ and $x=0.36$
since $\chi^s({\bm Q})$ develops toward $x\rightarrow 0$ and $0.4$. 
%For this reason, minimum value of $\lambda_{\rm E}$ is
%found to be realized at $x\sim0.1$.
Then, the double-dome behavior of $T_{\rm c}$ in LaFeAsO$_{1-x}$H$_x$ would be
explained, since $T_{\rm c}$ at both boundaries will be suppressed 
by the magnetic and orbital orders.
%add
At $x\approx0.4$,
the strongest spin fluctuations are mainly induced by 
the e-e FS nesting with ${\bm Q}\approx(\pi,\pi/3)$ 
shown in Fig. \ref{fig:Fermi} (c).
Using these fluctuations, $s_\pm$-wave like pairing is obtained since 
the $xy$-orbital hole-pockets exist even at $x=0.4$.
However, the SC gap of the electron-pockets becomes highly anisotropic 
%on the part of the FSs made of $d_{xz,yz}$-oritals
since the spin fluctuations of $xz/yz$-orbitals are very small.

Although the $d$-wave state is expected if the hole-pockets disappear
\cite{DHLee-Se,Scalapino-Se,Saito-RPA2}, the relation
$\lambda_{\rm E}(s_\pm$-wave$)>\lambda_{\rm E}(d$-wave$)$ 
is realized for $x\le0.4$ in the present study.
In both states, $\lambda_{\rm E}$ is quickly suppressed 
when the impurity concentration $n_{\rm imp}$ is finite,
meaning that the $s_{\pm}$- and $d$-wave states are 
fragile against impurities.

\begin{figure}[!htb]
\includegraphics[width=0.9\linewidth]{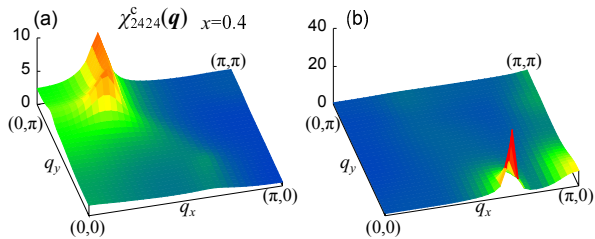}
\caption{
(Color online)
(a) $\chi^c_{2424}({\bm q})$ given by the RPA 
using the quadrupole interaction in Eq. (\ref{eqn:Hint})
for $\alpha_c=0.98$
(b) $\chi^c_{2424}({\bm q})$ given by the SC-VC method
using the Hubbard interaction for $\alpha_c=\alpha_s=0.98$.
}
\label{fig:chiQ}
\end{figure}

In the next stage, we study the development of the 
orbital susceptibility by $g$, by putting $U=0$.
Figure \ref{fig:Spp} (a) shows the $x$-dependence of $g_c$, 
given by the condition $\alpha_c=1$.
Similarly to Fig \ref{fig:Spm} (a), we introduce ${\bar g}(x)$ by liner
interpolation between $g_c$ for $x=0$ and that for $x=0.4$
in Fig. \ref{fig:Spp} (a).
Then, the orbital ordered states are realized for both $x\sim0$ and $x\sim0.4$.
The obtained $\chi^c_{2424}({\bm q})$ by the RPA for $x=0.4$ is shown in Fig. \ref{fig:chiQ} (a).
%Similar crossover from the e-h nesting to the e-e nesting with doping
%is also realized in the orbital susceptibility $\chi^c_{24,24}({\bm q})$.
%add
The strong orbital fluctuation at $x\sim0.4$ would originate from
the fact that the $t_{2g}$ orbitals degenerate \cite{Hosono-Hdoped},
and the relation $N_{xz}(0)=N_{yz}(0)\approx N_{xy}(0)$ at $x\sim0.4$.
%add
Experimental $T$-linear resistivity at $x\sim0.4$ 
would originate from the critical development of
orbital and spin fluctuations 
 \cite{Dagotto,Onari-rho}.

We also study the orbital-fluctuation-mediated $s_{++}$-wave state
for $g = {\bar g}(x)$:
Figure \ref{fig:Spp} (b) shows the obtained $\lambda_{\rm E}$ as a
function of $x$. 
%The behavior of $\lambda_{\rm E}$ in the $s_{++}$-wave state is similar to that
%of the $s_{\pm}$-wave state because orbital susceptibilities develop
%divergently toward $x\rightarrow 0$ and $0.4$.
Thus, double-dome behavior of $T_{\rm c}$ is 
explained by the orbital fluctuations.
The value of $\lambda_{\rm E}$ increases for larger cutoff energy $\omega_c$:
In Ref. \cite{Kontani-RPA}, we put $\omega_c=0.02$ since we considered the 
quadrupole interaction due to the Fe-ion oscillations.
However, $\omega_c$ for the effective quadrupole interaction due to VC
\cite{Onari-SCVC} depends on the electronic state and not unique.
In both cases, we should use larger $\omega_c$ since the 
used temperature ($T=0.02$) is much higher than the real $T_{\rm c}$.

%%%%%%%%%%%%%%%%%%%%%%%%%%%%%%%%%
%%%%%%%%%%%%%%%%%%%%%%%%%%%%%%%%%

Finally, we explain that the spin and orbital fluctuations 
mutually develop by taking the VC into account,
as actually observed in various Fe-based superconductors
\cite{Yoshizawa,Fujiwara}.
Beyond the RPA, $\hat{\chi}^0(q)$ in Eq. (\ref{eqn:RPA})
is replaced with $\hat{\chi}^0(q)+{\hat X}^{\rm s(c)}(q)$,
where ${\hat X}^{\rm s(c)}$ is the VC due to the Coulomb interaction
for the spin (charge) channel.
%In the RPA, the VC is neglected irrespective of its importance.
%The present authors found that the 
%quadrupole interaction in Eq. (\ref{eqn:Hint})
%is effectively induced by the VC due to Coulomb interaction,
%that is neglected in the RPA irrespective of its importance.
In Refs .\cite{Onari-SCVC,Ohno-SCVC,Tsuchiizu,Kontani-AL},
we have shown that 
%We developed the self-consistent VC (SC-VC) method in Ref .\cite{Onari-SCVC},
%and found that 
the Aslamazov-Larkin-type VC 
gives the large mode-coupling between spin and orbital, and then the
strong orbital fluctuations are triggered by the spin fluctuations.
This mode-coupling corresponds to the Kugel-Khomskii type 
spin-orbital coupling in the localized picture \cite{Devereaux}.

Figure \ref{fig:chiQ} (b) shows the strong development of 
antiferro-orbital fluctuations, $\chi^c_{2424}({\bm q})\gg1$, 
given by the SC-VC method for $x=0.4$, $U=1.1$, $J/U=0.073$ and $g=0$.
Therefore, the RPA analysis using ${\bar U}(x)$ and ${\bar g}(x)$ 
is justified by the SC-VC theory.
%In addition, strong ferro-orbital fluctuations, 
%that correspond to the orthorhombic structure transition,
%are also induced by the Aslamazov-Larkin-type VC for $x\sim0$ 
%\cite{Onari-SCVC,Kontani-AL}.
%The orbital fluctuations due to VC
%also occur in a simple two-orbital model,
%which would explain the nematic order in Sr$_3$Ru$_2$O$_7$
%\cite{Ohno-SCVC,Tsuchiizu}.
We have recently developed the SC-VC$\Sigma$ method, in which both the
self-energy and the VC are taken into account \cite{Onari-SCVC2}. 
Using this method, the $s_{++}$-wave state can be realized 
for realistic parameters $(J/U\sim0.1)$ for $x\sim0$ even for $g=0$. 
It is an important future problem to analyze the present
model using this method.

In summary, we have explained the reappearance of the 
spin and orbital orders in LaFeAsO$_{1-x}$H$_x$ at $x\sim0$ and $x\sim0.4$.
Both spin and orbital orders originate from the commensurate e-h FS nesting
for $x\sim0$, and incommensurate e-e FS nesting for $x\sim0.4$.
%Strong spin or orbital fluctuations stabilities at $\x\sim0$,
Due to strong spin and orbital fluctuations at $x\sim 0$ and $0.4$,
both the spin-fluctuation mediated $s_\pm$-wave state
and orbital-fluctuation mediated $s_{++}$-wave state 
can be realized, depending on the magnitude relation of these fluctuations.
Since small impurity effect on $T_{\rm c}$ for the first SC dome
\cite{Sato-imp,Li-1111} indicates the $s_{++}$-wave state,
%the $s_{++}\rightarrow$ nodal-s $\rightarrow s_\pm$ crossover 
the $s_{++}\rightarrow s_\pm$ crossover 
will occur with doping,
in case that the second SC dome is the $s_\pm$-wave state \cite{Kontani-RPA}.
Thus, the impurity effect study for $x\sim0.4$ is highly required.

%\acknowledgments
This study has been supported by Grants-in-Aid for Scientific 
Research from MEXT of Japan.
The part of Tokyo Tech was supported by JSPS First Program.
Numerical calculations were partially performed using 
the Yukawa Institute Computer Facility.

%%%%%%%%%%%%%%%%%%%%%%%%%
%references
%%%%%%%%%%%%%%%%%%%%%%%%

\end{document}